\documentclass[journal,comsoc]{IEEEtran}
\usepackage{amsmath,amssymb,amsfonts,float,graphicx,steinmetz,bm,subfigure,physics,mathtools,url,color,xcolor,etoolbox,ifthen,cite,dblfloatfix}
\usepackage{textcomp}
\usepackage{tikz}

\newcommand{\bi}{\begin{itemize}}
\newcommand{\ei}{\end{itemize}}

\newcommand{\be}{\begin{enumerate}}
\newcommand{\ee}{\end{enumerate}}
\newcommand{\bd}{\begin{description}}
\newcommand{\ed}{\end{description}}
\newcommand{\bc}{\begin{center}}
\newcommand{\ec}{\end{center}}
\newcommand{\bt}{\begin{tabbing}}
\newcommand{\et}{\end{tabbing}}
\newcommand{\bfig}{\begin{figure}}
\newcommand{\efig}{\end{figure}}
\newcommand{\beq}{\begin{equation}}
\newcommand{\beqarr}{\begin{eqnarray}}
\newcommand{\beqarrn}{\begin{eqnarray*}}
\newcommand{\eeq}{\end{equation}}
\newcommand{\eeqarr}{\end{eqnarray}}
\newcommand{\eeqarrn}{\end{eqnarray*}}
\newcommand{\bflr}{\begin{flushright}\vspace{-0.2in}}
\newcommand{\eflr}{\end{flushright}}
\newcommand{\bsub}{\begin{subequations}}
\newcommand{\esub}{\end{subequations}}
\newcommand{\barr}{\begin{array}}
\newcommand{\earr}{\end{array}}
\newcommand{\nn}{\nonumber}

\def\undb#1{\mbox{\bf{#1}}}

\begin{document}

\title{\huge{RIS-Assisted 6G Wireless Communications: A Novel Statistical Framework in the Presence of Direct Channel}
}

\author{Soumya~P.~Dash,~{\em Member, IEEE} and Aryan Kaushik,~{\em Member, IEEE}
\thanks{S. P. Dash is with the School of Electrical Sciences, Indian Institute of Technology Bhubaneswar, Argul, Khordha, Odisha 752050, India; e-mail: soumyapdashiitbbs@gmail.com.}
\thanks{A. Kaushik is with the School of Engineering and Informatics, University of Sussex, Brighton, UK, e-mail: aryan.kaushik@sussex.ac.uk.}
}
\maketitle
\begin{abstract}
A RIS-assisted wireless communication system in the presence of a direct communication path between the transceiver pair is considered in this paper. The transmitter-RIS and the RIS-receiver channels follow independent Nakagami-$m$ distributions, and the direct channel between the transceiver pair follows a Rayleigh distribution. Considering this system model, the statistics of the composite channel for the RIS-assisted communication system are derived in terms of obtaining novel expressions for the probability density functions for the magnitude and the phase of the communication channel. The correctness of the analytical framework is verified via Monte Carlo simulations, and the effects of the shape parameters of the channels and the number of reflecting elements in the RIS on the randomness of the composite channel are studied via numerical results.
\end{abstract}
%
\section{Introduction}
The development of sixth-generation (6G) wireless communication networks is progressing since telecommunication industries and standardization bodies are working towards realizing the full potential of future 6G wireless standards \cite{dashTII2022}. The International Telecommunication Union (ITU-R) enlists several emerging technology trends for the development of IMT-2030 (6G), including reconfigurable intelligent surfaces (RIS) \cite{renzoPROC2022}. The RIS technology leverages smart radio surfaces with a high number of small antennas or metamaterial elements based on a programmable structure that can be used to control the propagation of electromagnetic (EM) waves \cite{dashrisCL2022, miguelDCN2022, basuVTC2023, shlezingerWCM2021}. At higher frequency spectrums such as fifth generation (5G) new radio (NR) frequencies/millimeter wave (mmWave), sub-terahertz (THz), and THz, there is a gain of large bandwidth and uncongested bands; however, these frequencies are prone to high path loss and blockage effects leading to supporting short-range communication. However, with the use of RIS technology, a strong and controllable non-line-of-sight (NLoS) wireless communication link can be formed.

The reflection of EM waves makes passive RIS elements highly compatible with the transceiver antenna setup, which can provide high capacity and coverage while benefitting from its energy-efficient characteristics. RIS configurations can be realized by incorporating a large number of antenna elements with reconfigurable processing networks, which provides a continuous antenna aperture. RIS technology becomes essential for network reliability and extreme connectivity in the next generation of wireless standards. Furthermore, while passive RIS does not practically consume any direct-current power due to no amplification of the dynamically steered signal, active RIS employs an active reflection-type amplifier to amplify the reflected signal \cite{zhangTCOM2023}. Utilizing RIS has also been expanded to several key technology enablers for 6G, such as joint sensing and communications (JSAC) \cite{vlachosWCNC2023} and non-terrestrial networks \cite{tokaSPAWC2023}. The RIS elements can be intelligently reconfigured using index modulation, which is crucial for tuning the directional precoded signal \cite{singh2023, basu2023}.

Like any communication system, a statistical framework for a RIS-assisted system is quintessential to studying its performance and proposing novel solutions for such systems. Various studies have been reported in the literature which derive statistical distributions of wireless channels with respect to the use of RIS systems. The authors in \cite{triguiARX2020} have studied the performance of a RIS-assisted wireless system under various fading environments to obtain expressions for ergodic capacity and outage probability. The authors in \cite{badiuWCL2020} have employed the central limit theorem (CLT) to model the composite channel of a RIS system with large number of reflecting elements to follow a Nakagami-$m$ distribution. The integral expression for the probability density function (p.d.f.) of the fading channel for a RIS system assisting the communication for a user with a single antenna at the receiver is proposed in \cite{ferreiraOJC2020}. The performance of a RIS-assisted system with two different phase configurations of the RIS elements in terms of using integral expressions and CLT to model the statistics of the communication channel is studied by the authors in \cite{selimisCL2021}. Furthermore, the authors in \cite{aratiWCL2022} and \cite{dashCL2022} consider the effect of correlation amongst the communication channels to derive the statistical characteristics of a RIS-assisted system. The p.d.f. of the composite channel for a RIS-assisted system with the individual channels following Nakagami-$m$ distributions is derived in \cite{TeTy:22}.

In all of the above-mentioned studies, the effect of the direct channel between the transceiver pair hasn't been considered. However, the importance of such a direct link towards the performance of RIS-assisted communications has been shown in \cite{huangPROC2022}-\cite{lyuWCL2020}. Furthermore, such scenarios are practically observed in various vehicular, small-cell, and Internet of Things applications in indoor and outdoor channel scenarios. However, a majority of work carried out to obtain the statistics of such channels rely on using several approximations, including CLT, which does not justify the effect of the actual composite RIS-assisted channel towards the performance of such systems. Thus, to fill in this research gap, we consider a RIS-assisted wireless communication system with a direct channel present between the transceiver pair. Considering the channels for the transmitter-RIS and the RIS-receiver pairs to follow independent Nakagami-$m$ distributions and the direct channel between the transceiver pair to follow a Rayleigh distribution, the exact statistics of the composite channel are derived in terms of obtaining series-form expressions for the joint p.d.f.s of the real and imaginary parts of the composite channel. Using this, the magnitude and the phase distributions of the composite channel are also obtained, which are verified by Monte Carlo simulations for varying channel parameters. Numerical results corroborating the analytical framework reveal that the shape parameters and the number of RIS elements have contrary effects on the randomness of the magnitude and the phase distributions of the communication channel for the RIS-assisted system.

The rest of the paper is organized as follows. Section II describes the system model of the RIS-assisted system in the presence of the direct communication link. The statistics of the modified channel in terms of obtaining the above-mentioned p.d.f.s using the first principles without using any approximations are derived in Section III. This is followed by the numerical results elaborating the analytical framework in Section IV and the concluding remarks in Section V.
\section{RIS-assisted System Model with Direct Communication Channel}
\begin{figure}[htp]
\centering
\includegraphics[width=3.4in,height=1.6in]{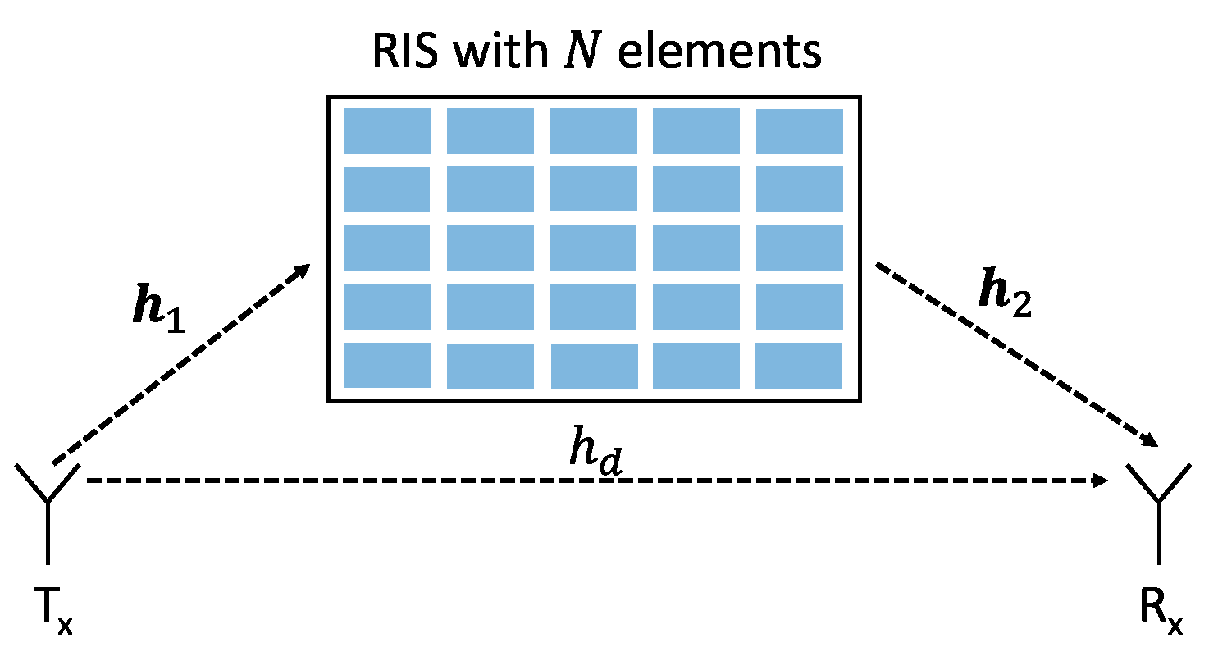}\vspace{-0.3cm}
\caption{System model of RIS-assisted wireless system with a direct communication channel.}
\label{f1}
\noindent
\end{figure}

The model of the RIS-assisted wireless communication system is depicted in Fig. \ref{f1}.
The transmitter and the receiver are considered to communicate using one antenna each, and the data transmission occurs via two communication links, namely the direct link between the transmitter and the receiver and the link by utilizing the RIS consisting of $N$ reconfigurable meta-surfaces. Denoting the transmitted symbol by $s$, the received symbol is thus given as
\beq
r = \left( \undb{h}_1^T \mathbf{\Phi} \undb{h}_2 + h_d \right) s + n \, ,
\label{eq1}
\eeq
where $\undb{h}_1$ and $\undb{h}_2$ are the $N \times 1$ channel gains between the transmitter and the RIS and the RIS and the receiver, respectively, and $h_d$ is the direct channel between the transceiver pair. $\left( \cdot \right)^T$ denotes the transpose operator, and $\mathbf{\Phi}$ is the $N \times N$ diagonal phase-shift matrix whose diagonal elements are consisted of the phase shifts introduced by the RIS, i.e., $\mathbf{\Phi} = \text{diag}\left( \phi_1 ,\ldots, \phi_N \right)$. Furthermore, $n$ is the additive noise in the wireless channel which follows a zero mean complex Gaussian distribution implying that $n \sim {\mathcal{CN}} \left( 0, \sigma_n^2 \right)$. The envelopes of the channel gains $h_{1,k}$s and $h_{2,k}$s for $k=1,\ldots,N$ are considered to follow Nakagami-$m_1$ and Nakagami-$m_2$ distributions, respectively, and the direct channel follows a zero mean Complex Gaussian distribution implying that $h_d \sim {\mathcal{CN}} \left( 0, \sigma_h^2 \right)$. Moreover, all the channel gains involved are considered to be statistically independent of each other. From (\ref{eq1}), we denote the modified channel as
\beqarr
h \! \! \! \! &=& \! \! \! \! \undb{h}_1^T \mathbf{\Phi} \undb{h}_2 + h_d \nn \\
&=& \! \! \! \! \sum_{k=1}^N \left| h_{1,k} \right| \left| h_{2,k} \right|
\exp \left\{ \angle h_{1,k} + \angle h_{2,k} + \phi_k \right\} + h_d \, ,
\label{eq2}
\eeqarr
where $\left| \cdot \right|$ and $\angle \cdot$ denote the magnitude and phase operators, respectively, for a complex variable. For maximizing the signal-to-noise ratio at the receiver end and to maximize the performance of the system, the phase shifts of the RIS are selected as $\phi_k = - \left( \angle h_{1,k} + \angle h_{2,k} \right)$, which results in the modified channel as
\beq
h = \sum_{k=1}^N \left| h_{1,k} \right| \left| h_{2,k} \right| + h_d \, .
\label{eq3}
\eeq
\section{Statistics of the Modified Channel}
In this section, we compute the statistics of the modified channel coefficient given in (\ref{eq3}). It is to be noted that the modified channel coefficient $h$ is a complex random variable. Furthermore, owing to the statistical independence of the actual channel gains and $h_d$ being a circular Gaussian random variable, the real and imaginary parts of $h$, denoted by $h_r$ and $h_i$, respectively, become statistically independent. Thus, the p.d.f. of the modified channel coefficient is given by
\beq
f \left( h \right) = f_{h_r,h_i} \left( x , y \right)
= f_{h_r} \left( x \right) f_{h_i} \left( y \right) \, ,
\label{eq4}
\eeq
where, from (\ref{eq3}),
\beqarr
h_r \! \! \! \! &=& \! \! \! \!
\sum_{k=1}^N \left| h_{1,k} \right| \left| h_{2,k} \right|
+ \Re \left\{ h_d \right\} 
= h_{r,1} + h_{r,2} \, , \nn \\
h_i \! \! \! \! &=& \! \! \! \! \Im \left\{ h_d \right\} \, ,
\label{eq5}
\eeqarr
with $\Re \left\{ \cdot \right\}$ and $\Im \left\{ \cdot \right\}$ denoting the real part and imaginary part operators, respectively.
\subsection{Statistics of $h_r$}
From (\ref{eq5}), we have $h_{r,2} = \Re \left\{ h_d \right\}$ and from the statistics of $h_d$, we have the p.d.f. of $h_{r,2}$ to be given as
\beq
f_{h_{r,2}} \left( \tilde{x} \right) = \frac{1}{\sqrt{\pi \sigma_h^2}}
\exp \left\{ - \frac{\tilde{x}^2}{\sigma_h^2} \right\} \, ,
\tilde{x} \in \left( -\infty , \infty \right) \, .
\label{eq6}
\eeq
Owing to the statistical independence of $h_{r,1}$ and $h_{r,2}$, the p.d.f. of $h_r$ can be computed as
\beqarr
f_{h_r} \left( x \right) \! \! \! \! &=& \! \! \! \!
\int_{-\infty}^{\infty} f_{h_{r,1}} \left( \tilde{x} \right)
f_{h_{r,2}} \left( x - \tilde{x} \right) \text{d} \tilde{x} \nn \\
&=& \! \! \! \! \frac{1}{\sqrt{\pi \sigma_h^2}} \int_{-\infty}^{\infty}
\exp \left\{ - \frac{ \left( x - \tilde{x} \right)^2}{\sigma_h^2} \right\}
f_{h_{r,1}} \text{d} \tilde{x} \, .
\label{eq7}
\eeqarr
Using the expansion of the exponential function and the binomial expansion, we have
\beqarr
\exp \left\{ - \frac{ \left( x - \tilde{x} \right)^2}{\sigma_h^2} \right\}
\! \! \! \! &=& \! \! \! \! \sum_{\ell=0}^{\infty}
\frac{\left( -1\right)^{\ell} \left(x-\tilde{x} \right)^{2 \ell}}
{\sigma_h^{2\ell} \ell!} \nn \\
&=& \! \! \! \! \sum_{\ell=0}^{\infty} \sum_{p=0}^{2 \ell}
\frac{\left( -1 \right)^{3\ell-p} \left(2\ell \right)! x^p}
{\sigma_h^{2 \ell} \ell! p! \left( 2\ell-p\right)!} \tilde{x}^{2 \ell-p} .
\label{eq8}
\eeqarr
Substituting (\ref{eq8}) in (\ref{eq7}), we have
\beq
f_{h_r} = \frac{1}{\sqrt{\pi \sigma_h^2}}
\sum_{\ell=0}^{\infty} \sum_{p=0}^{2 \ell}
\frac{\left( -1 \right)^{3\ell-p} \left(2\ell \right)! x^p}
{\sigma_h^{2 \ell} \ell! p! \left( 2\ell-p\right)!}
\int_{-\infty}^{\infty} \tilde{x}^{2\ell-p} f_{h_{r,1}} \text{d} \tilde{x}.
\label{eq9}
\eeq
It can be noted from (\ref{eq9}) that the integral is equivalent to the $\left(2\ell-p\right)$-th moment of $h_{r,1}$. Furthermore, from the statistics of $\undb{h}_1$ and $\undb{h}_2$, the random variable $h_{r,1}$ is equivalent to the sum of double-Nakagami-m random vectors and thus, using \cite[Eq: 11]{TeTy:22} followed by algebraic simplifications leads to the expression of the p.d.f. of $h_r$ to be obtained as
\beqarr
f_{h_r} \left( x \right) \! \! \! \! &=& \! \! \! \!
\frac{1}{\sqrt{\pi \sigma_h^2}}
\sum_{\ell=0}^{\infty} \sum_{p=0}^{2 \ell}
\frac{\left( -1\right)^{3 \ell-p} \left( 2 \ell \right)!
\left( \Omega_1 \Omega_2 \right)^{\ell-\frac{p}{2}} x^p}
{\sigma_h^{2\ell} \left( m_1 m_2 \right)^{\ell-\frac{p}{2}}
\ell! p! \left(2\ell - p \right)!} \nn \\
&& \times \sum_{i_1=0}^{m_1-1} \cdots \sum_{i_N=0}^{m_1-1} \prod_{k=1}^N
\frac{\left(m_2 \right)_{m_1-1-i_k} \left( 1-m_2 \right)_{i_k}}
{\left( m_1-1 \right)! i_k! \Gamma \left( u \right)} \nn \\
&& \times \Gamma \left( \ell +1 - \frac{p}{2} \right)
\Gamma \left( \ell + u - \frac{p}{2} \right) \, , x \in \left( -\infty , \infty \right) \nn \\
\label{eq10}
\eeqarr
where $u=N \left( m_1+m_2-1 \right) - \sum_{k=1}^N i_k$, $\Gamma \left( \cdot \right)$ denotes the Gamma function, and $\left( i \right)_k$ is the Pochhammer symbol. Furthermore, $\undb{E} \left[ \left| h_{1,k} \right|^2 \right] = \Omega_1$ and $\undb{E} \left[ \left| h_{2,k} \right|^2 \right] = \Omega_2$, with $\undb{E} \left[ \cdot \right]$ denoting the expectation operator.
\subsection{Statistics of $h_i$}
From (\ref{eq5}) and the statistics of $h_d$, we have $h_i \sim {\mathcal{N}} \left( 0 , \sigma_h^2/2 \right)$. Thus, the p.d.f. of $h_i$ is expressed as
\beq
f_{h_i} \left( y \right) = \frac{1}{\sqrt{\pi \sigma_h^2}}
\exp \left\{ - \frac{y^2}{\sigma_h^2} \right\} \, ,
y \in \left( -\infty , \infty \right) \, .
\label{eq11}
\eeq
\subsection{Statistics of the channel}
Substituting the results obtained in (\ref{eq10}) and (\ref{eq11}) to (\ref{eq4}), the p.d.f. of $h$, equivalent to the joint p.d.f. of $h_r$ and $h_i$ is obtained as
\beqarr
&& \! \! \! \! \! \! \! \! \! \! \! \!
f \left( h \right) = f_{h_r,h_i} \left( x , y \right) \nn \\
&& \! \! \! \! \! \! \! \!
= \frac{1}{\pi \sigma_h^2} \exp \left\{ -\frac{y^2}{\sigma_h^2} \right\}
\sum_{\ell=0}^{\infty} \sum_{p=0}^{2 \ell}
\frac{\left( -1\right)^{3 \ell-p} \left( 2 \ell \right)!
\left( \Omega_1 \Omega_2 \right)^{\ell-\frac{p}{2}} x^p}
{\sigma_h^{2\ell} \left( m_1 m_2 \right)^{\ell-\frac{p}{2}}
\ell! p! \left(2\ell - p \right)!} \nn \\
&& \! \! \! \! \! \!
\times \sum_{i_1=0}^{m_1-1} \cdots \sum_{i_N=0}^{m_1-1} \prod_{k=1}^N
\frac{\left(m_2 \right)_{m_1-1-i_k} \left( 1-m_2 \right)_{i_k}}
{\left( m_1-1 \right)! i_k! \Gamma \left( u \right)} \nn \\
&& \! \! \! \! \! \!
\times \Gamma \left( \ell +1 - \frac{p}{2} \right)
\Gamma \left( \ell + u - \frac{p}{2} \right) \, ,
\left\{ x,y \right\} \in \left( -\infty , \infty \right) .
\label{eq12}
\eeqarr

To compute the envelope of phase distributions of $h$, we utilize the method of change of variables, which would result in the joint distribution of the envelope and phase of $h$ to be given as
\beqarr
&& \! \! \! \! \! \! \! \! \! \! \! \!
f_{\left| h \right|, \angle h} \left( r, \theta \right)
= r f_{h_r,h_i} \left( x , y \right)
\big|_{x=r \cos \theta , y = r\sin \theta} \nn \\
&& \! \! \! \! \! \! \! \!
= \frac{r}{\pi \sigma_h^2} \exp \left\{ -\frac{r^2 \sin^2 \theta}
{\sigma_h^2} \right\}
\sum_{\ell=0}^{\infty} \sum_{p=0}^{2 \ell}
\frac{\left( -1\right)^{3 \ell-p} 
\left( \Omega_1 \Omega_2 \right)^{\ell-\frac{p}{2}}}
{\sigma_h^{2\ell} \left( m_1 m_2 \right)^{\ell-\frac{p}{2}}} \nn \\
&& \! \! \! \! \! \!
\times \frac{\left( 2 \ell \right)! r^p \cos^p \theta}
{\ell! p! \left(2\ell - p \right)!}
\sum_{i_1=0}^{m_1-1} \! \! \cdots \! \! \sum_{i_N=0}^{m_1-1} \prod_{k=1}^N
\frac{\left(m_2 \right)_{m_1-1-i_k} \left( 1-m_2 \right)_{i_k}}
{\left( m_1-1 \right)! i_k! \Gamma \left( u \right)} \nn \\
&& \! \! \! \! \! \!
\times \Gamma \left( \ell +1 - \frac{p}{2} \right)
\Gamma \left( \ell + u - \frac{p}{2} \right) \, ,
r \geq 0, \theta \in \left( -\pi , \pi \right) .
\label{eq13}
\eeqarr
Thus, the p.d.f. of the envelope of the modified channel gain can be obtained as
\beq
f_{\left| h \right|} \left( r \right) = \int_{-\pi}^{\pi}
f_{\left| h \right|, \angle h} \left( r, \theta \right) \text{d} \theta \, .
\label{eq14}
\eeq
Upon substituting (\ref{eq13}) in (\ref{eq14}), we need to solve the integral given as
\beq
I_r = \int_{-\pi}^{\pi} \cos^p \theta
\exp \left\{-\frac{r^2 \sin^2 \theta}{\sigma_h^2} \right\} \text{d} \theta \, .
\label{eq15}
\eeq
It is to be noted that the value of the integration in (\ref{eq15}) equals zero for odd values of $p$. Thus, the integration value exists only for even values of $p$, which results in the expression of $I_r$ to be obtained as
\begin{small}
\beqarr
I_r \! \! \! \! &=& \! \! \! \! \int_{-\pi}^{\pi} \cos^{2p} \theta
\exp \left\{-\frac{r^2 \sin^2 \theta}{\sigma_h^2} \right\} \text{d} \theta \nn \\
&\stackrel{(a)}{=}& \! \! \! \! \sum_{q=0}^{\infty}
\frac{\left( -1\right)^q r^{2q}}{\sigma_h^{2q}}
\int_{\pi}^{\pi} \cos^{2p} \theta \sin^{2q} \theta \text{d} \theta \nn \\
&=& \! \! \! \! \sum_{q=0}^{\infty}
\frac{\left( -1\right)^q 2^q r^{2q}}{\sigma_h^{2q}}
\Gamma \left( q + \frac{1}{2} \right)
{}_2\tilde{F}_1 \left( -2p, q + \frac{1}{2}; 2q+1 ; 2 \right) \, , \nn \\
\label{eq16}
\eeqarr
\end{small}
\hspace{-0.15cm}where the step $(a)$ is obtained by using the expansion of the exponential function and ${}_2\tilde{F}_1 \left(\cdot, \cdot ; \cdot ; \cdot \right)$ denotes the regularized ${}_2F_1$ hypergeometric function. Substituting (\ref{eq16}) in (\ref{eq15}) and (\ref{eq14}) followed by algebraic simplifications results in the final expression for the p.d.f. of $\left| h \right|$ to be given as
\beqarr
&& \! \! \! \! \! \! \! \! \! \! \! \! \! \! \! \!
f_{\left| h \right|} \left( r \right) \nn \\
&& \! \! \! \! \! \! \! \! \! \! \! \!
= \sum_{\ell=0}^{\infty} \sum_{p=0}^{\infty} \sum_{q=0}^{\infty}
\frac{\left(-1 \right)^{3\ell-2p+q} 2^q
\Gamma \left( q+\frac{1}{2}\right) \left(2 \ell \right)!}
{\pi \sigma_h^{2\ell+2q+1} \ell! \left( 2p\right)!
\left( 2\ell-2p\right)!} \nn \\
&\times& \! \! \! \!
\frac{\left(\Omega_1 \Omega_2 \right)^{\ell-p}}{\left(m_1 m_2 \right)^{\ell-p}}
r^{2p+2q+1} {}_2\tilde{F}_1 \! \!
\left( -2p, q + \frac{1}{2}; 2q+1 ; 2 \right) \nn \\
&\times& \! \! \! \!
\sum_{i_1=0}^{m_1-1} \! \! \cdots \! \! \sum_{i_N=0}^{m_1-1} \prod_{k=1}^N
\frac{\left(m_2 \right)_{m_1-1-i_k} \left( 1-m_2 \right)_{i_k}}
{\left( m_1-1 \right)! i_k! \Gamma \left( u \right)} \nn \\
&& \qquad \quad \times \Gamma \left( \ell +1 - p \right)
\Gamma \left( \ell + u - p \right) \, , r \geq 0.
\label{eq17}
\eeqarr

Similarly, to compute the phase distribution of the modified channel coefficient, we need to solve the expression given as
\beq
f_{\angle h } \left( \theta \right) = \int_{0}^{\infty}
f_{\left| h \right|, \angle h} \left( r, \theta \right) \text{d} r \, .
\label{eq18}
\eeq
Upon substituting (\ref{eq13}) in (\ref{eq18}) we encounter an integral which is expressed as
\beq
I_{\theta} = \int_{0}^{\infty} r^{p+1}
\exp \left\{ - \frac{r^2 \sin^2 \theta}{\sigma_h^2} \right\} \text{d} r
=\frac{\sigma_h^{p+2} \Gamma \left( \frac{p}{2}+1 \right)}
{2 \sin^{p+2} \theta} \, .
\label{eq19}
\eeq
Utilizing (\ref{eq19}) and (\ref{eq17}) in (\ref{eq18}) followed by algebraic simplifications results in the expression for the p.d.f. of the phase of the channel coefficient as
\begin{small}
\beqarr
f_{\angle h } \left( \theta \right) \! \! \! \! &=& \! \! \! \!
\frac{1}{\pi} \sum_{\ell=0}^{\infty} \sum_{p=0}^{\infty}
\frac{\left( -1\right)^{3\ell-p} \left( 2 \ell \right)!
\Gamma \left( \frac{p}{2} + 1\right) \cot^p \theta
\left( \Omega_1 \Omega_2 \right)^{\ell-\frac{p}{2}}}
{2 \sigma_h^{2\ell} \ell! p! \left( 2 \ell -p\right)!
\left( m_1 m_2 \right)^{\ell-\frac{p}{2}}} \nn \\
&& \ \times
\sum_{i_1=0}^{m_1-1} \! \! \cdots \! \! \sum_{i_N=0}^{m_1-1} \prod_{k=1}^N
\frac{\left(m_2 \right)_{m_1-1-i_k} \left( 1-m_2 \right)_{i_k}}
{\left( m_1-1 \right)! i_k! \Gamma \left( u \right)} \nn \\
&& \ \times \Gamma \left( \ell +1 - \frac{p}{2} \right)
\Gamma \left( \ell + u - \frac{p}{2} \right) \, , \, 
\theta \in \left(-\pi , \pi \right] .
\label{eq20}
\eeqarr
\end{small}
\section{Numerical Results}
The numerical results corroborating the analysis carried out in this paper are presented in this section. Specifically, we present the computation and simulation plots of the various p.d.f.s derived. For all the plots, we consider $\Omega_1=1$, $\Omega_2=1$, and $\sigma_h^2=1$, and vary the Nakagami-$m$ shape parameters and the number of RIS elements.

\begin{figure}[htp]
\centering
\includegraphics[width=3.4in,height=2.6in]{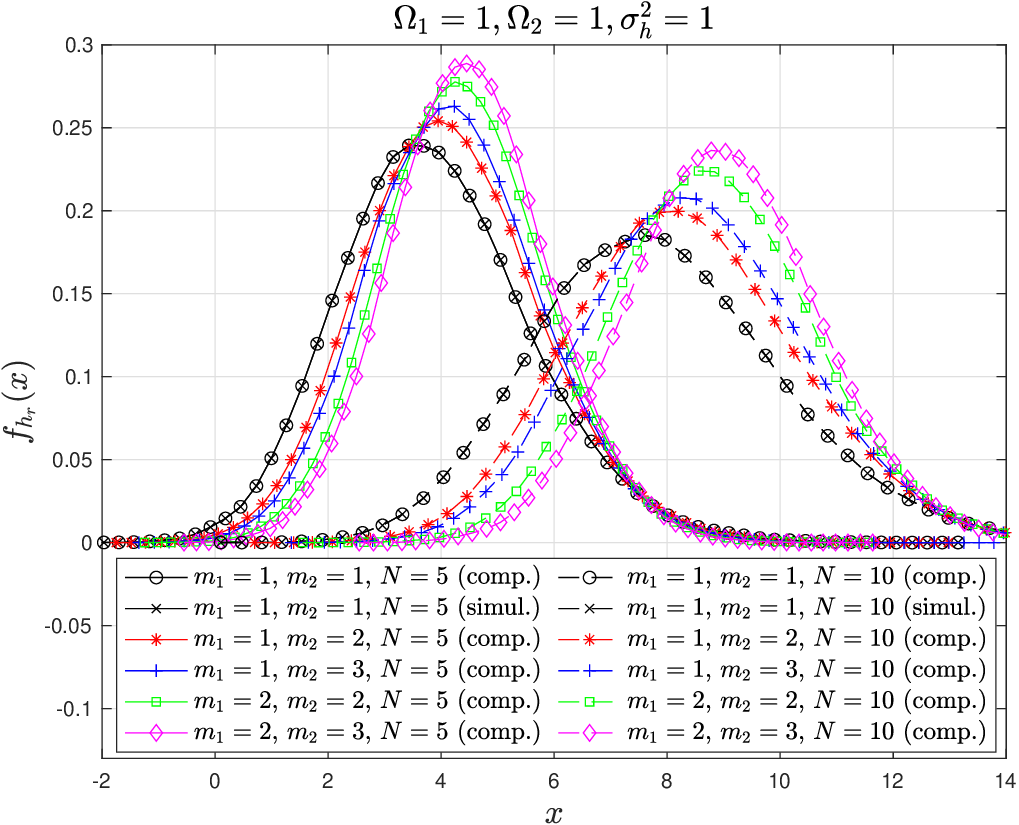}\vspace{-0.3cm}
\caption{$f_{h_r} (x)$ versus $x$ for $\Omega_1=\Omega_2=\sigma_h^2=1$, $m_1=1,2$, $m_2=1,2,3$, and $N=5,10$.}
\label{f2}
\noindent
\end{figure}
Fig. \ref{f2} presents the plots of p.d.f. of the real part of the modified channel, wherein the simulation plots are generated for $10^8$ Monte Carlo iterations and the computation plots are obtained from (\ref{eq10}) by considering a sufficiently large number of terms in the summations leading to the relative error to be less than 0.0001\%. The exactness of the plots justifies the correctness of the analytical framework. Furthermore, it is observed that the peak values of the p.d.f.s increase with an increase in the values of the shape parameters of either of the channels and with decreasing values of $N$. Moreover, the mean values of the p.d.f.s increase with increasing values of $m_1$, $m_2$, and $N$. Thus, the modified channel of the RIS system becomes more random for lower values of the shape parameters and higher values of the RIS elements.

\begin{figure}[htp]
\centering
\includegraphics[width=3.4in,height=2.6in]{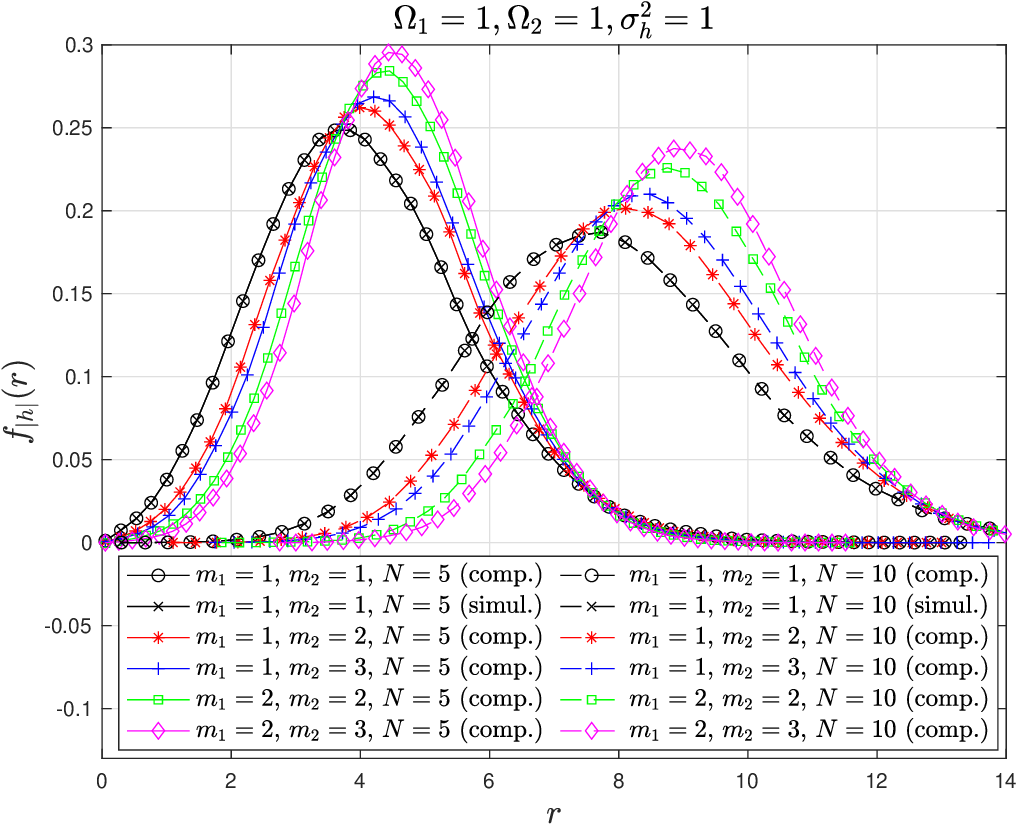}\vspace{-0.3cm}
\caption{$f_{|h|} (r)$ versus $r$ for $\Omega_1=\Omega_2=\sigma_h^2=1$, $m_1=1,2$, $m_2=1,2,3$, and $N=5,10$.}
\label{f3}
\noindent
\end{figure}
The plots for the p.d.f. of the magnitude of the channel gain with the computation plots being generated using (\ref{eq17}) are presented in Fig. \ref{f3}. The computation and the simulation plots match, justifying the correctness of the analysis. Further, similar observations as Fig. \ref{f2} are also observed here. Thus, the randomness of the channel magnitude reduces with an increase in the values of $m_1$ and $m_2$ and for lower values of RIS elements. These show practically the effect of the RIS elements on controlling the communication system.


\begin{figure}[ht] 
\centering
\subfigure[$N=5$]{
\includegraphics[width=3.4in,height=2.6in]{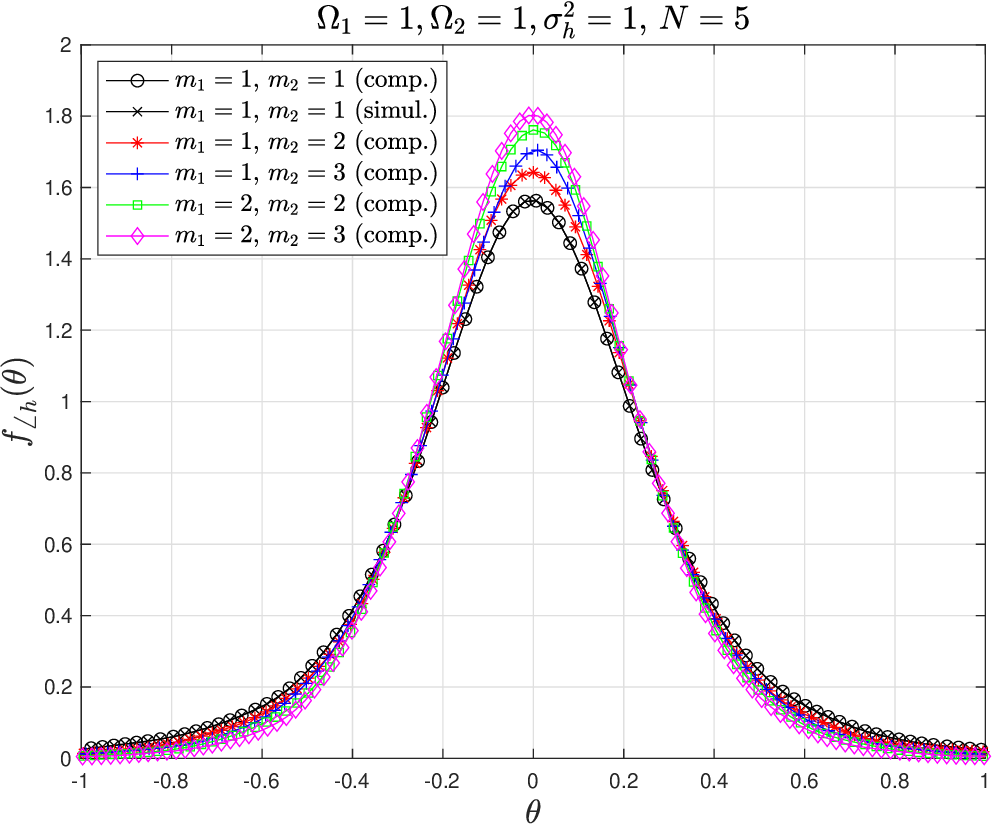}%
\label{f4a}
}\hfill
\subfigure[$N=10$]{
\includegraphics[width=3.4in,height=2.6in]{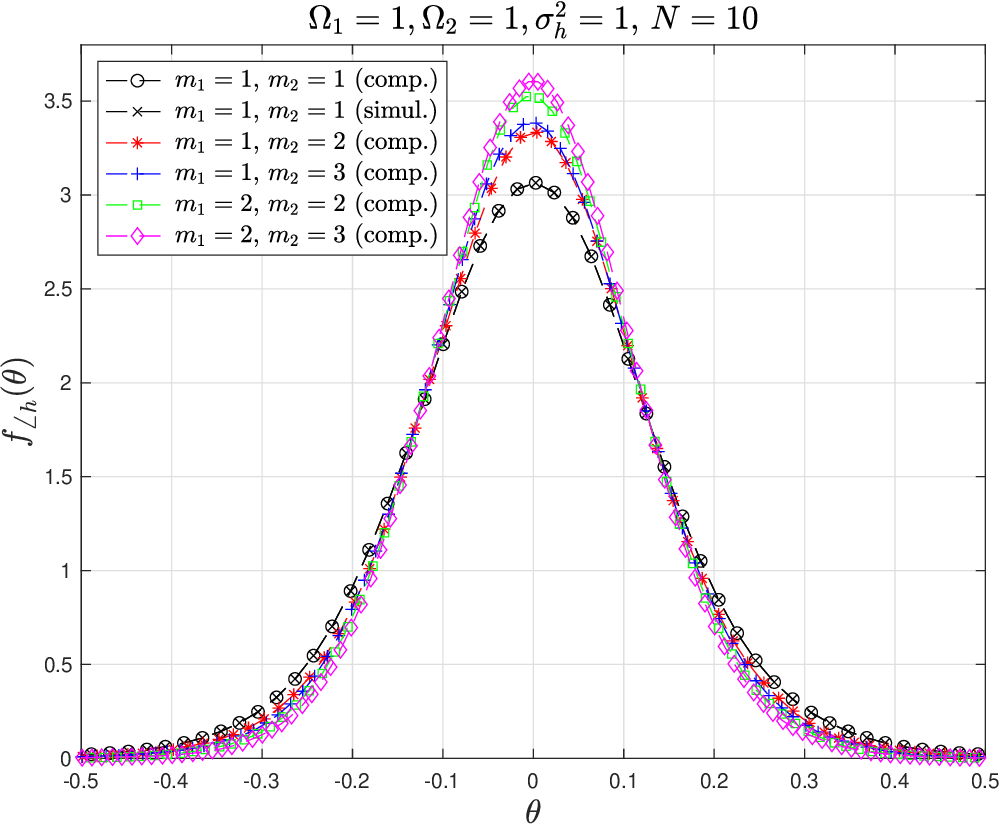}%
\label{f4b}
} \caption{$f_{\angle h} \left(\theta \right)$ versus $\theta$ for $\Omega_1=\Omega_2=\sigma_h^2=1$, $m_1=1,2$, $m_2=1,2,3$, and (a) $N=5$ (b) $N=10$.}
\end{figure}

Figs. \ref{f4a} and \ref{f4b} present the plots of the phase of the modified channel of the system for $N=5$ and $N=10$, respectively. Similar to previous results, the computation and simulation results match, thus justifying the correctness of the analysis. Furthermore, it is observed that the phase distribution is even and doesn't follow a uniform distribution. Rather, the shape of the phase distribution resembles that of a Gaussian curve with a mean value of zero. Moreover, the spread of the distribution (equivalent to the random variable's variance) becomes smaller with an increasing number of reflecting elements in the RIS and the values of $m_1$ and $m_2$. Thus, increasing the number of RIS elements reduces the randomness of the phase distribution of the modified RIS channel.
\section{Conclusion}
We consider a RIS-assisted wireless communication system where the data communication between the transmitter and receiver occurs via the RIS and a direct channel between the transceiver pair. Considering the communication channels via the RIS elements to follow independent Nakagami-$m$ distributions and the direct channel to follow a Rayleigh distribution, the statistics of the composite channel are derived using the first principles. Specifically, novel expressions for the p.d.f.s of the magnitude and phase of the composite channel are obtained, which are corroborated via numerical studies. The practical significance of the RIS modifying the physical communication channel is observed in the sense of the randomness of the magnitude and the phase components of the composite channel increasing and reducing, respectively, with an increase in the value of $N$.

\end{document}